\documentclass[acus]{pac99}

\usepackage{epsfig}

\newcommand{\bit}{\begin{Itemize}}
\newcommand{\eit}{\end{Itemize}}

\begin{document}
\title{
POTENTIAL HAZARDS FROM NEUTRINO RADIATION AT MUON COLLIDERS
}
\author{ Bruce J. King, Brookhaven National Laboratory $^1$ }
\maketitle

\begin{abstract}

  High energy muon colliders, such as the TeV-scale conceptual designs now
being considered, are found to produce enough high energy neutrinos to
constitute a potentially serious off-site
radiation hazard in the neighbourhood of the accelerator site.
A general characterization of this radiation hazard is given,
followed by an order-of-magnitude calculation for the off-site annual
radiation dose and a discussion of accelerator design and site
selection strategies to minimize the radiation hazard.

\end{abstract}

\footnotetext[1]{
web page: http://pubweb.bnl.gov/people/bking/,
email: bking@bnl.gov.
This work was performed under the auspices of
the U.S. Department of Energy under contract no. DE-AC02-98CH10886. }

\section{Introduction}
\label{sec-intro}

 Current conceptual designs for muon colliders~[1]
envisage large currents of opposing positively and negatively charged
muons circulating in a collider ring until decay into neutrinos and electrons:
\begin{eqnarray}
\mu^- & \rightarrow & \nu_\mu + \overline{\nu_{\rm e}} + {\rm e}^-,
                                             \nonumber \\
\mu^+ & \rightarrow & \overline{\nu_\mu} + \nu_{\rm e} + {\rm e}^+.
                                                 \label{nuprod}
\end{eqnarray}

This will produce an intense disk of neutrinos emanating out in the plane of
the collider ring. The vertical angular divergence of the neutrino disk
can be characterized by the spread in the
relative angle between the neutrino and muon directions and, from
relativistic kinematics, the neutrinos in the forward hemisphere
in the muon rest frame are boosted, in the laboratory frame, into
a narrow cone with an opening half-angle,
\begin{equation}
\theta_\nu \simeq \sin \theta_\nu = 1/\gamma =
\frac{m_\mu c^2}{E_\mu} \simeq \frac{10^{-4}}{E_\mu [{\rm TeV}]},
                                                   \label{thetanu}
\end{equation}
with $\gamma$ the relativistic boost factor of the muon,
$E_\mu$ the muon energy and $m_\mu$ the muon rest mass.

  The vertical angular spread of the neutrino disk could, in principle,
also receive contributions from the angular spread of the muon beam. However,
for reasonable magnet lattice designs this will usually
produce negligible additional divergence everywhere around the collider
ring except at the final focus regions around collider experiments.

  The potential radiation hazard comes from the showers of ionizing
particles produced in interactions
of neutrinos in the soil and other objects bathed by
the disk. The tiny interaction cross-section for neutrinos is
greatly compensated by the huge number of high energy neutrinos produced
at muon colliders.

\section{Quantitative Derivation of Radiation Dose}

  A quantitative expression for the radiation dose received by
a person from the decay of  $N_\mu$ muons of each sign, at tangential
distance $L$ from an idealized perfectly circular muon collider ring
and in the plane of the ring is given by:
\begin{eqnarray}
D^{ave}  = \frac{N_\mu}{L^2}
\int_{0}^{2\pi} \frac{d\theta}{2\pi}
      \cdot \frac{1}{4\pi} \: \frac{d\Omega'}{d\Omega}(\theta) \times
                 \nonumber \\
\sum_{\nu type, i=1,4} \int_{0}^{1} dx f^{i} (x)
\sigma^{i}(E_\nu) d^i(E_\nu),
       \label{doseexpress}
\end{eqnarray}
where ${\rm E_\nu}$ is a function of both integration variables, $x$
and $\theta$ and the variables and form of the expression are explained
in the following paragraphs.

  In principle, neutrinos can be emitted in all directions relative
to the muon trajectory at decay so
the angle between the muon beam
and the neutrino direction, $\theta$, is formally integrated over
all muon directions around ring, 
However, equation~\ref{nuprod} shows that most of the
contribution to the radiation dose will come from neutrinos oriented
within of order $1/\gamma$ or less from the muon beam direction, so the
size of the collider ring can be ignored.
Clearly, a fraction
$\frac{d\theta}{2\pi}$ of the muons will decay in the angular increment
$d\theta$ and, for the reasonable assumption that the muon beams are
unpolarized on average, the neutrino decays will be isotropic in muon
rest frame and the fraction of neutrino decays per unit solid angle in the
laboratory rest frame is
is $\frac{1}{4\pi} \frac{d \Omega'}{d \Omega}(\theta)$,
where primed coordinates denote the muon rest frame and unprimed coordinates
the laboratory rest frame. 
  Relativistic kinematics gives:
\begin{equation}
\frac{d \Omega'}{d \Omega} = \gamma  (1 + \beta \cos \theta')
                       \frac{d\theta'}{d\theta},
\end{equation}
$\gamma = 9.46 \times 10^3 \times E_\mu [TeV]$ and
$\beta \simeq 1$ for realistic collider energies.
This allows the integration over the laboratory angle, $\theta$
in equation~\ref{doseexpress} to be replaced by an integration over the angle
in the muon rest frame, $\theta'$.

 A biological target in the radiation disk is tangent to the collider ring
in two places and so will receive neutrinos from the decays of both positive
and negative muons.  Therefore, the neutrino type index, i, runs over all
4 neutrino types -- $\overline{\nu_{\rm e}}$, $\nu_\mu$,
$\overline{\nu_\mu}$ and $\nu_{\rm e}$.

The energy probability distribution in the muon rest frame for the
production of neutrino type i is $dx \cdot f_{x}^{i}$, with $x$ defined
as the fraction of the maximum neutrino energy in the muon
rest frame: $E'_\nu = \frac{x.m_\mu c^2}{2}$. The explicit form for
$f$ is known to be:
$f = 6.x^2 - 4.x^3$ for unpolarized muon-type neutrinos or anti-neutrinos
and $f = 12.x^2 - 12.x^3$ for unpolarized electron-type neutrinos or
anti-neutrinos. Boosting to the laboratory frame gives

\begin{equation}
E_\nu(x,\cos \theta') = x \cdot \frac{E_\mu}{2}
                  \left( 1 + \beta \cos \theta' \right).
                           \label{boost}
\end{equation}

 The cross-section per nucleon, $\sigma^{i}(E_\nu)$, is  expressed, for
now, in the same units of length as L and $d^i(E_\nu)$ is the average
radiation dose from a neutrino of type i and energy $E_\nu$ interacting
in each nucleon of a biological target, assuming the equilibrium
approximation and expressed in the same units
as $D^{ave}$.

  Most of the ionization energy dose deposited
in a person will come from interactions in the soil and other objects in the
person's vicinity rather than from the more direct process of neutrinos
interacting inside a person. At TeV energy scales, much less than one
percent of the energy flux from the daughters
of such interactions will be absorbed in the relatively small amount of matter
contained in a person, with the rest passing beyond the person.

 Equation~\ref{doseexpress} implicitly assumes the
simplifying ``equilibrium approximation''
that the ionization energy absorbed by a person is equal to the energy of
the showers initiated by interactions in that person.

 It seems reasonable to assume that the equilibrium approximation should
give either a reasonable estimate or a conservative overestimate of the
radiation dose absorbed by a person for most of the innumerable possible
distributions of matter.  From conservation of energy,
it would clearly be a good approximation for
the reference case of a homogeneous medium of any density sufficient that
the radial extent of the hadronic and electromagnetic showers initiated by
neutrino interactions is small compared to the height of the neutrino
radiation disk. In realistic geometries, some of the shower energy will
typically leak
out to beyond the extent of the neutrino disk through low density regions
of air etc., presumably decreasing the radiation dose to below the
equilibrium estimate.

The radiation dose in units of Sieverts (Sv)
is numerically almost equal to the energy deposition in a biological
target in units of J/kg
for the energetic hadronic and electromagnetic showers from neutrino
interactions. Therefore,
if $k^i$ is defined to be the weighting factor for converting
from neutrino energy to radiation dose for a neutrino of type i
then $k^i$ will be
numerically equal to the average fraction of the interaction energy ending
up as electrons or hadrons.

 The radiation dose per neutrino of energy $E_\nu$ is then given by
$d^i(E_\nu)[Sv] \simeq 1.6 \times 10^{-7} \times k^i \times E_\nu[TeV]
\times (10^3 \times N_{Avogadro})$,
where $1.6 \times 10^{-7}$ is the conversion factor from Joules to TeV
and the numerical factor $10^3 \times N_{Avogadro}$ arises because
Sieverts are defined as energy absorbed per kilogram rather than
per nucleon.

 Because neutrino cross-sections are almost linear with energy in the TeV
range they can be expressed as
$\sigma^{i}(E_\nu)[cm^2] \simeq \sigma^{i}_R[cm^2/TeV] \times E_\nu [TeV]$,
where the ``reduced cross section'', $\sigma^{i}_R$, can be approximated
as a constant over a fairly large energy range.

 On making these substitutions and integrating over the muon angle,
equation~\ref{doseexpress} can be rewritten as:
\begin{eqnarray}
D^{ave} [Sv] = 4.5 \times 10^{15} \times
 \frac{N_\mu \times (E_\mu[TeV])^3}{(L[km])^2} \times
                 \nonumber \\
\sum_{\nu type, i=1,4} \sigma_R^{i}[cm^2/TeV] \times k^i
\times \int_{0}^{1} dx f^{i}(x) \cdot x^2,
                                   \label{doseworking}
\end{eqnarray}
where the units of all dimensioned variables are
given in square brackets.

 The dominant interaction processes of TeV-scale neutrinos are charged
current (CC) and neutral current (NC) deep inelastic scattering off nucleons:
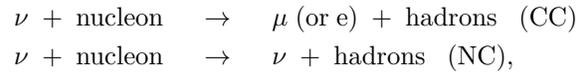
\begin{eqnarray*}
\nu\;+\;{\rm nucleon}\;\; & \rightarrow & \;\;
                 \mu \; ({\rm or\; e})\; +\; {\rm hadrons}\;\;\;({\rm
                                               CC})  \\
\nu\;+\;{\rm nucleon}\;\; & \rightarrow & \;\;
                                \nu \;+\; {\rm hadrons}\;\;\;({\rm NC}),
                                                \label{nudis}
\end{eqnarray*}
 It is the subsequent interactions of the daughter hadrons and electrons,
initiating showers of ionizing particles, that constitute the bulk of
the radiation hazard. In contrast, that part of the neutrino energy
transferred into daughter neutrinos or muons will almost all be transported
to outside the radiation disk rather than contributing to the energy
absorbtion of a person within the disk.

 The numerical
calculations for equation~\ref{doseworking} are summarized in table 1.
Substituting this into equation~\ref{doseworking} gives:
\begin{equation}
D^{ave} [Sv] = 3.7 \times 10^{-23} \times
 \frac{N_\mu \times (E_\mu[TeV])^3}{(L[km])^2}.
                           \label{dosevalue}
\end{equation}

\begin{table}[htb!]
\caption{Contributions to the radiation dose from the different
types of neutrino interactions.  The reduced cross-secttion,
$\sigma_R^i$, is specified for
100 GeV neutrinos and using a simple model for the nucleon
in which the cross-section ratio for neutrinos to anti-neutrinos
was assumed to be 2:1 and ignoring the small differences between
the average hadronic fractions for NC and CC interactions.
The reduced cross-section and
product are in units of ${\rm 10^{-35} cm^2/TeV}$.}
\begin{tabular}{|ccccc|}
\hline
int., $i$ & $\sigma^i_R$ & $k^i$
  &  $<x^2>^i$  & $\sigma^i_R \cdot k^i \cdot <x^2>^i$  \\
\hline
$\nu_\mu-CC$ & 0.722 & 0.458 & 0.533 & 0.176 \\
$\nu_\mu-NC$ & 0.226 & 0.458 & 0.533 & 0.055 \\
$\nu_{\rm e}-CC$ & 0.722 & 1.000 & 0.400 & 0.289 \\
$\nu_{\rm e}-NC$ & 0.226 & 0.458 & 0.400 & 0.041 \\
$\overline{\nu_\mu}-CC$ & 0.375 & 0.292 & 0.533 & 0.058 \\
$\overline{\nu_\mu}-NC$ & 0.131 & 0.292 & 0.533 & 0.020 \\
$\overline{\nu_{\rm e}}-CC$ & 0.375 & 1.000 & 0.400 & 0.150 \\
$\overline{\nu_{\rm e}}-NC$ & 0.131 & 0.292 & 0.400 & 0.015 \\
\hline
SUM  & & & & 0.804 \\
\hline
\end{tabular}
\label{table: contributions}
\end{table}

  It will now be shown that
the radiation intensity would be expected to vary greatly around the
neutrino disk, depending on the detailed design of the collider ring magnet
lattice, so the value of $D_{ave}$ by itself is not sufficient to assess
the radiation hazard for any particular collider design. For example,
it is clear from the derivation of equation~\ref{dosevalue} that
the radiation contribution tangent to the collider ring at a dipole
bending magnet will be proportional to the beam's bending radius
at the magnet, which is inversely proportional to the magnetic field strength.

 For even bigger variations,
tangents to the collider ring at anywhere other than a dipole
magnet the muon currents will travel in straight line trajectories and
the neutrinos will line up as local radiation ``hot spots'' in the
radiation disk
-- cones of more intense radiation with characteristic opening half-angles
of $\theta_\nu = 1/\gamma$.

 The contribution from straight sections is given by an equation
analagous to equation~\ref{doseexpress}:
\begin{eqnarray}
D^{ss}  = \frac{f^{ss} \times N_\mu}{L^2}
\times  \frac{\gamma^2}{\pi} \times
                 \nonumber \\
\sum_{\nu type, i=1,2} \int_{0}^{1} dx f^{i} (x)
\sigma^{i}(E_\nu) d^i(E_\nu),
                    \label{ssexpress}
\end{eqnarray}
where $f^{ss}$ is the length of the straight section as a fraction of the
collider circumference
\begin{equation}
f^{ss} = L/C
\end{equation}
 and the factor $\gamma^2 / \pi$ is
the fraction of neutrinos decaying in the forward direction per unit solid
angle after being boosted from isotropic decays in the muon rest frame
into the laboratory frame.

 The summation in equation~\ref{ssexpress} is now only over the 2
neutrino types
produced by the sign of muon travelling in the considered direction,
i.e., either $\overline{\nu_{\rm e}}$ and $\nu_\mu$ for $\mu^-$ decays or 
$\overline{\nu_\mu}$ and $\nu_{\rm e}$ for $\mu^+$ decays
(equation~\ref{nuprod}).
 The summed contributions in table 1
for $\mu^+$ and $\mu^-$ are very nearly equal, so it
is reasonable to use the average contribution,
$0.402 \times {\rm 10^{-35} cm^2/TeV}$, for either sign.

  Following a similar derivation to that for equation~\ref{dosevalue}
the numerical value for the dose is:
\begin{equation}
D^{ss} [Sv] = 1.1 \times 10^{-18} \times
 \frac{f^{ss} \times N_\mu \times (E_\mu[TeV])^4}{(L[km])^2}.
                                \label{ssvalue}
\end{equation}

 The radiation cones from the final focus regions
around collider experiments are important exceptions to
equation~\ref{ssvalue}, since
the muon beam itself will have an angular divergence in these regions that
may be large enough to spread out the neutrino beam by at least an order of
magnitude in both x and y.

 More detailed calculations to check and refine these calculations,
using Monte Carlo-based particle tracking computer simulations, are
in progress.

  On comparing equations~\ref{dosevalue} and~\ref{ssvalue} it is
easily seen that the length of straight section to produce an extra
contribution equal to the
planar average dose, ${\rm l_{equiv}}$, is approximately:
\begin{equation}
l_{equiv} [meters] \simeq 0.034 \times \frac{C[km]}{E_\mu[TeV]}
                      \simeq \frac{0.71}{B_{ave}[T]},
                                    \label{lequiv}
\end{equation}
where the final expression uses the relation between muon energy,
ring circumference and and average bending magnetic field in units
of Tesla:
\begin{equation}
C[km] = \frac{2 \pi \cdot E_\mu[TeV]}{0.3 \cdot B_{ave}[T]},
                                    \label{circum}
\end{equation}
valid for a circular ring.

 Two mitigating factors come into play at many-TeV energies to
reduce the radiation rise with energy:
\begin{enumerate}
  \item  the neutrino cross section begins to rise significantly less
rapidly than linearly with neutrino energy
  \item  the radiation disk (or cone) becomes so narrow that the
``equilibrium approximation'' is no longer accurate because much of
the induced shower of charged particles leaks out transversely
beyond the extent of the radiation disk. The ``cut-off'' width
at which the equilibrium approximation will begin to fail badly
is the typical transverse shower size in whatever medium is
initiating the showers. This will be of order a meter for
typical solids. (As an aside, it will be hundreds of meters in air
so the equilibrium approximation will be overly conservative at
all collider energies
for people in open areas rather than surrounded by massive objects.)

\end{enumerate}

\section{Strategies to Minimize Off-Site Radiation Doses}

 Because of the strong dependence on muon energy, the radiation levels
rapidly become a serious design constraint for colliders at the TeV
scale and above. For illustration, table 2 gives the predicted radiation
levels for some example muon collider parameter sets~[3]. 
 For comparison, the U.S.
federal off-site radiation limit is $10^{-3}$ Sv/year, which is of the
same order of
magnitude as the typical background radiation from natural causes (i.e.
0.4 to $4 \times 10^{-3}$ Sv/yr~[2]) and it is assumed that
acceptable radiation levels must be considerably lower than these
values.

  As a desirable design strategy for all energies, it is clear that
great care must be taken
to minimize or eliminate long straight sections in the collider ring.
For example, the magnet lattice could consist partly or entirely of
dual function magnets,
where the beam focusing and bend are accomplished in the same magnets.
Optionally, it might be convenient to retain one or two long straight
sections by constructing radiation enclosures around where their
radiation hot spots exit the ground.

 Perhaps the most direct way of decreasing the radiation levels
is to greatly decrease the muon current. This can be done either
by sacrificing luminosity (as in the 4 TeV parameter set of
table 2) or, more attractively, by increasing the luminosity
per given current through better muon cooling or other technological
advances.

 Further, one might consider placing the accelerator deep underground
so the radiation disk won't reach the surface for some distance.
For the example of a very flat region of the Earth the exit
distance to the surface ${\rm L_{exit}}$ will be related to
the collider depth, $D$, and the Earth's radius,
$R_E = 6.4 \times 10^6 {\rm m}$, by
$L_{exit} = \left( 2 \times D \times R_E \right) ^{1/2}$,
where the three parameters are in consistent units of length, e.g., meters.
Substituting into equations~\ref{dosevalue} and~\ref{ssvalue} gives

\begin{equation}
D^{ave}_{exit} [Sv] = 2.9 \times 10^{-24} \times
 \frac{N_\mu \times (E_\mu[TeV])^3}{(D[m])}
                           \label{davexit}
\end{equation}
and
\begin{equation}
D^{ss}_{exit} [Sv] = 4.1 \times 10^{-24} \times
\frac{l[m] \times B_{ave} \times N_\mu \times (E_\mu[TeV])^3}{(D[m])},
                           \label{dssexit}
\end{equation}
respectively, where equations~\ref{lequiv} and~\ref{circum} have been
substituted into the second of these equations.

 It is seen that the radiation dose at exit falls inversely with
collider depth. The quadratic dependence of the depth on
$L_{exit}$ means that exit distances of order 10 km are easily
achievable, but achieving an $L_{exit}$ of order 100 km is already
starting to require a prohibitively large depth.

 Further speculative options that have been discussed include
(i) tilting the ring to take best advantage of the local topography,
(ii) placing the collider ring on a hill so the radiation disk
passes harmlessly above the surroundings 
and, even more speculatively,
(iii) spreading out and diluting the
neutrino radiation disk
by continuously sweeping the muon beam
orbit in a vertical plane using dipole corrector
magnets.

 Even when the preceding strategies have been used,
the strong rise in neutrino energy probably dictates that
muon colliders at CoM energies of beyond a few TeV will probably have to be
constructed at isolated sites where the public would not be exposed to
the neutrino radiation disk at all. This would definitely be required
for the 10 TeV and 100 TeV parameter sets of table 2.
Because of the additional costs this would
involve, these will presumably be ``second generation'' machines, arriving
after the technology of muon colliders has been established in one or more
smaller and less expensive machines built at existing HEP laboratories.


 In conclusion, some order-of-magnitude calculations have been presented
which show that the neutrino-induced radiation hazard might be a very
serious problem for high energy muon colliders.
The neutrino radiation problem appears to impose severe constraints
on the site selection for a muon collider complex and on the layout
of the complex.

 It is speculated that the highest energy muon (and hadron) colliders
and their associated neutrino radiation disks may be required to be
enclosed within a huge new world HEP laboratory somewhere where there
is a large area of cheap, sparsely populated land.

\section{Acknowledgements}

  This paper has benefitted greatly from discussions with collaborators
in the muon collider collaboration, particularly Dr. Robert
Palmer and Dr. Nikolai Mokhov.


\section{references}


\noindent [1] The Muon Collider Collaboration,
``Status of Muon Collider Research
and Development and Future Plans'', to be submitted to Phys. Rev. E.

\noindent [2]     The Particle Data Group,
    Review of Particle Physics,
    Phys. Rev. D54 (1996).

\noindent [3] B.J. King,
 ``Discussion on Muon Collider Parameters at Center of Mass
Energies from 0.1 TeV to 100 TeV'', 19 June, 1998, 
     Submitted to Proc. Sixth European Particle Accelerator
Conference (EPAC'98), Stockholm, Sweden, 22-26 June, 1998.
Available at http://pubweb.bnl.gov/people/bking/.

\newpage

\begin{table}[htb!]
\begin{center}
\caption{Radiation dose estimates for the example muon collider
parameter sets of reference~[3].}

\begin{tabular}{|r|ccccc|}
\hline
\multicolumn{1}{|c|}{ {\bf center of mass energy, ${\rm E_{CoM}}$} }
                            & 0.1 TeV & 1 TeV & 4 TeV &  10 TeV  & 100 TeV \\
\multicolumn{1}{|c|}{ {\bf description} }
                            & MCC para. set & LHC complement & E frontier &
                                        $2^{\rm nd}$ gen. & ult. E scale \\
luminosity, ${\cal L}$ [${\rm cm^{-2}.s^{-1}}$]
                                        & $1.2 \times 10^{32}$
                                        & $1.0 \times 10^{34}$
                                        & $6.2 \times 10^{33}$
                                        & $1.0 \times 10^{36}$
                                        & $4.0 \times 10^{36}$ \\
\hline \hline
\multicolumn{1}{|l|}{\bf relevant collider parameters:} & & & &  & \\
circumference, C [km]                   & 0.3 & 2.0 & 7.0 & 15 & 100 \\
ave. bending B field [T]               & 3.5 & 5.2 & 6.0 & 7.0 & 10.5 \\
($\mu^-$ or) $\mu^+$/bunch,${\rm N_0[10^{12}}]$
                                        & 4.0 & 3.5 & 3.1 & 2.4 & 0.18 \\
($\mu^-$ or) $\mu^+$ bunch rep. rate, ${\rm f_b}$ [Hz]
                                        & 15 & 15 & 0.67 & 15 & 60 \\
ave. current [mA]                      & 20 & 10 & 0.46 & 24 & 4.2 \\
beam power [MW]                        & 1.0 & 8.4 & 1.3 & 58 & 170 \\
time to beam dump,
          ${\rm t_D} [\gamma \tau_\mu]$ & no dump & 0.5 & 0.5 & no dump & 0.5 \\
effective turns/bunch                  & 519 & 493 & 563 & 1039 & 985 \\
\hline
\hline
\multicolumn{1}{|l|}{\bf neutrino radiation parameters:} & & & & & \\
collider reference depth, D[m]           & 10 & 125 & 300 & 300 & 300 \\
$\nu$ beam distance to surface [km]    & 11 & 40 & 62 & 62 & 62 \\
$\nu$ beam radius at surface [m]       & 24 & 8.4 & 3.3 & 1.3 & 0.13 \\
str. sect. length for 10x ave. rad.,
${\rm L_{x10}}$[m] & 1.9 & 1.3 & 1.1 & 1.0 & 2.4 \\
ave. rad. dose in plane [mSv/yr]               & $3 \times 10^{-5}$
                                        & $9 \times 10^{-4}$
                                        & $9 \times 10^{-4}$
                                        & 0.66 & 6.7 \\
 \hline

\end{tabular}
\label{specs}
\end{center}
\end{table}
 
\end{document}